\documentclass[letter]{aa} 
\usepackage{natbib}
\bibpunct{(}{)}{;}{a}{}{,}
\usepackage{graphicx}
\usepackage{txfonts}
\begin{document}
   \title{Tracing the sites of obscured star formation in the Antennae galaxies 
          with {\em Herschel}-PACS
          \thanks{{\em Herschel} is an ESA space observatory with science instruments
                  provided by European-led Principal Investigator consortia
                  and with important participation from NASA.}
          }


   \author{U.~Klaas
          \inst{1}
          \and
           M.~Nielbock
          \inst{1}
          \and
           M.~Haas
          \inst{2}
          \and
           O.~Krause
          \inst{1}
          \and
           J.~Schreiber
          \inst{1}
          }

   \institute{Max-Planck-Institut f\"ur Astronomie (MPIA),
              K\"onigstuhl 17, 69117 Heidelberg, Germany \\
              \email{klaas@mpia.de}
         \and
             Astronomisches Institut, Ruhr-Universit\"at Bochum (AIRUB), 
              Universit\"atsstr. 150/NA7, 44801 Bochum, Germany \\
             }

   \date{Received 31 March 2010 / Accepted 19 April 2010}

 
  \abstract
   {}
   {FIR imaging of interacting galaxies allows locating even hidden sites
    of star formation and measuring of the relative strength of nuclear and
    extra-nuclear star formation. We want to resolve the star-forming sites in 
    the nearby system of the Antennae.}
   {Thanks to the unprecedented sharpness and depth of the PACS camera onboard 
    ESA's {{\em Herschel}} Space Observatory, it is possible for the first time to 
    achieve a complete assessment of individual star-forming knots in the FIR 
    with scan maps at 70, 100, and 160\,$\mu$m. We used clump extraction 
    photometry and SED diagnostics to derive the properties related to 
    star-forming activity.}
   {The PACS 70, 100, and 160\,$\mu$m maps trace the knotty structure of 
    the most recent star formation along an arc between the two nuclei in the 
    overlap area. The resolution of the starburst knots and additional 
    multi-wavelength data allow their individual star formation history 
    and state to be analysed. In particular, the brightest knot in the 
    mid-infrared (K1), east of the southern nucleus, exhibits the highest 
    activity by far in terms of dust heating and star formation rate, 
    efficiency, and density. With only 2\,kpc in diameter, this area has a 
    10--1000\,$\mu$m luminosity, which is as high as that of our Milky Way. 
    It shows the highest deficiency in radio emission in the radio-to-FIR 
    luminosity ratio and a lack of X-ray emission, classifying it as a very 
    young complex. The brightest 100 and 160\,$\mu$m emission region (K2), 
    which is close to the collision front and consists of 3 knots, also shows 
    a high star formation density and efficiency and lack of X-ray emission in 
    its most obscured part, but an excess in the radio-to-FIR luminosity
    ratio. This suggests a young stage, too, but different conditions in its 
    interstellar medium. Our results provide important checkpoints for 
    numerical simulations of interacting galaxies when modelling the star 
    formation and stellar feedback.}
   {}


   \maketitle
%

\section{Introduction}

We have set up a photometric imaging programme of a sample of nearby 
interacting galaxies as part of the {\em Herschel} Guaranteed Time Key Project 
SHINING\footnote{http://www.mpe.mpg.de/ir/Research/SHINING/} to trace
and resolve the youngest star formation sites that are still embedded in dust,
making them strong far-infrared (FIR) emitters. The \object{Antennae} 
(\object{Arp~244}, \object{NGC~4038}/\object{NGC~4039}) is a key object in 
this sample, since it is an interactive system, where strong extranuclear star 
formation was found previously in the so-called overlap area of the two disks 
between the two nuclei. Optical images show dark dust lanes interspersed with 
a few HII regions in this area \citep[e.g.][]{whitmore95}. It houses several 
supergiant molecular cloud complexes 
\citep{stanford90,wilson00,gao01,schulz07}, which provide a large 
reservoir for bursts of star formation. Prominent 6 and 20\,cm radio 
emission \citep{hummel86}, which spatially matches the CO emisison peaks, 
suggests there are ongoing starbursts. Compressed magnetic fields are found 
towards the northern edge indicating pre-starbursts \citep{chyzy04}. 

Mid-infrared (MIR) observations have discovered embedded starbursts at the 
southeastern border of the overlap region \citep{vigroux96,mirabel98,wilson00},
while submm 850 $\mu$m observations suggest a large amount of cooler dust in 
the northern overlap region, arguing in favour of relatively less powerful 
starbursts there \citep{haas00,zhu03,schulz07}.

In the case of buried starbursts, one expects to see the continuum
re-emission of the hiding dust at mid- and far-infrared wavelengths.
However, previous FIR space missions have not reached the spatial resolution
required to separate the overlap region from the nuclei. {\em Herschel}-PACS now 
offers the capability of such spatially resolved observations for the first 
time. 

\section{Observations and data reduction}

\object{Arp~244} has been observed with the Photodetector Array Camera and
Spectrograph (PACS) \citep{poglitsch10} onboard the {\em Herschel} Space Observatory 
\citep{pilbratt10} on Operational Day 208, between 2009-12-08T02:13:51Z and 
2009-12-08T03:01:26Z as part of the Science Demonstration Phase (SDP) using 
the three photometric bands centred on nominal wavelengths of 70, 100, and 
160 $\mu$m. Four scan maps, 2 each in two orthogonal directions (45 and 
135\,deg), have been obtained with the medium scan speed of 
$20\arcsec$\,s$^{-1}$. The design of the scans and cross-scans resulted in 8 
scan legs of $6\arcmin$ length and 10 scan legs of $4\farcm5$ length, 
respectively. The leg separation was $39\arcsec$.

   \begin{figure*}[ht!]
   \begin{center}
   \includegraphics[angle=-90,width=0.88\textwidth]{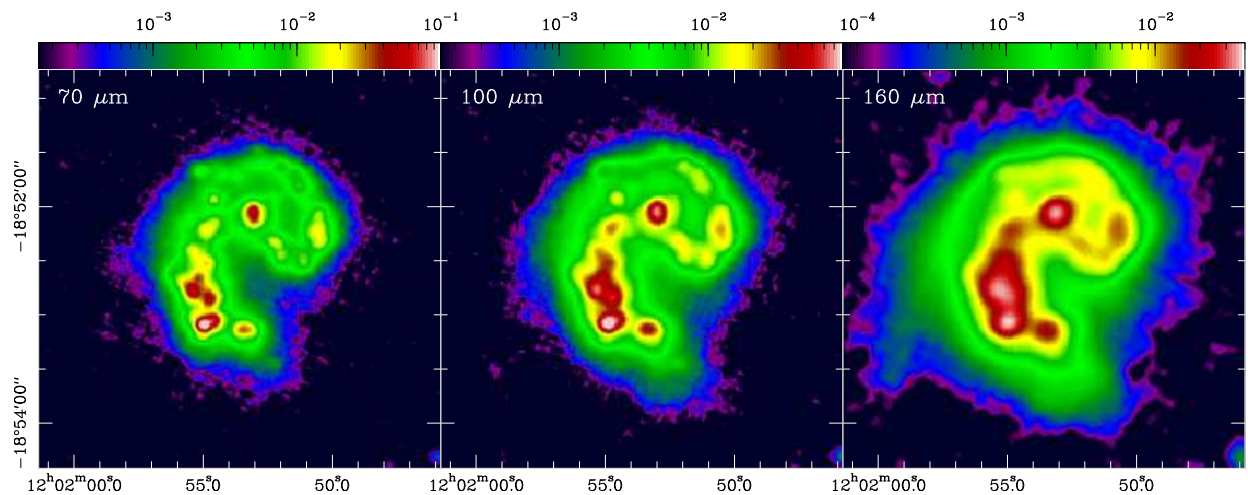}
   \caption{PACS maps of the Antennae galaxies at 70, 100 and
         160\,$\mu$m. The look-up tables indicate the flux level in Jy/$\Box"$.}
   \label{pacsimages}
   \end{center}
   \end{figure*}

The raw data were reduced with the {\em Herschel} Interactive Processing
Environment (HIPE)\footnote{HIPE, as part of the {\em Herschel} Common Science System
(HCSS), is a joint development by the {\em Herschel} Science Ground Segment
Consortium, consisting of ESA, the NASA {\em Herschel} Science Center, and the HIFI,
PACS, and SPIRE consortia. (See
\texttt{http://herschel.esac.esa.int/DpHipeContributors.shtml})} version 3.0,
build 1134, running the standard steps of the PACS photometer pipeline
\citep{poglitsch10} and leading to level 1 calibrated data. Deglitching of the 
data was done via a moderate setting of the MMT deglitching task and a 
second-level deglitching on the intermediate level 2 products. A source mask 
was designed covering the emission area in the map that allowed the emission 
features to be excluded from the deglitching and the subsequent high-pass 
filtering of the data. It was applied with a median window of 41 data samples 
to remove the effect of bolometer signal drifts and the $1/f$ noise. The flux 
correction factors provided by the PACS Instrument Control Centre (ICC) team 
have been applied. The FWHM of the PSF is 5\farcs5, 6\farcs8, and 11\farcs3 in 
the 70, 100, and 160\,$\mu$m band, respectively. Astrometry of the PACS maps 
was corrected on three compact field sources.

For 15 $\mu$m photometry, the ISOCAM map \citep{mirabel98}, provided to us by 
O.~Laurent for our earlier publication \citep{haas00}, was used.
For 24 $\mu$m photometry, the {\em Spitzer}-MIPS scan-map from programme 32 
(P.I. G. Fazio) was reduced using the Data Analysis Tool 
\citep[DAT;][]{gordon05}. 

The photometry of the partially crowded emission knots of the 15\,$\mu$m ISOCAM,
24\,$\mu$m {\em Spitzer}-MIPS and 70, 100, and 160\,$\mu$m {\em Herschel}-PACS maps used 
the HIIphot package developed by D.~Thilker \citep{thilker00}. The standard 
parameter settings were applied, except a maximum source size of 2000\,pc and 
a background annulus of 1500\,pc were used in order to have similar source 
borders, taking the variation in spatial resolution into account over the 
considered wavelength range. PSF FWHM and the termgrad array were adjusted 
according to the instrument/wavelength and the flux level of the individual 
maps. 

\section{Results}

\begin{figure}[ht!]
   \begin{center}
   \includegraphics[width=0.79\columnwidth]{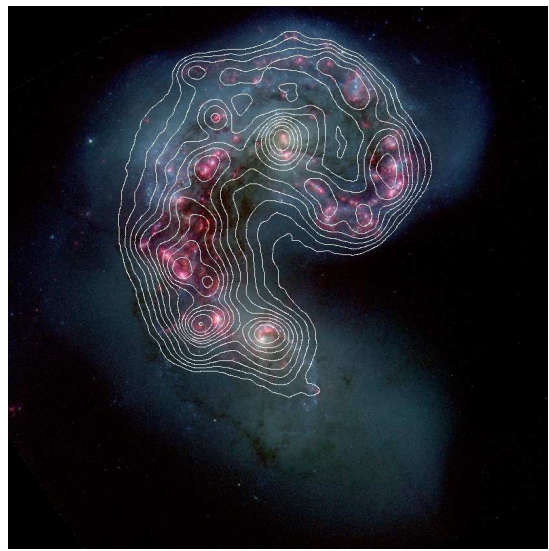}
   \caption{PACS 100\,$\mu$m contours superimposed on a composite of HST-images 
            in the F435W (blue), F550M (green), and F658N (H$_\alpha$) filters.}
   \label{pacs100contoursonHSTBVHalpha}
   \end{center}
   \end{figure}

Figure~\ref{pacsimages} shows the {\em Herschel}-PACS maps at 70, 100, and 160 $\mu$m.
The 160\,$\mu$m map reaches farthest out, since it has twice the exposure time
of the blue (70$+$100) bands. Its faintest contours cover the optical disks 
(cf.\ Fig.~\ref{pacs100contoursonHSTBVHalpha}). There is some dust emission 
excess, where the optical image shows the deep cleft between the two disks. The 
160\,$\mu$m image also indicates the base of the southern tidal arm where it 
shows a bulge in the optical, but there is no emission detected from the long 
thin tidal arms. The 70 and 100\,$\mu$m maps show a cut-off southwest of the 
NGC\,4039 nucleus, and this very low flux level indicates a lack of heating 
sources and possibly also a depletion of dust in this area 
(Fig.~\ref{pacs100contoursonHSTBVHalpha}).

   \begin{figure}[ht!]
   \begin{center}
   \includegraphics[angle=-90,width=0.55\columnwidth]{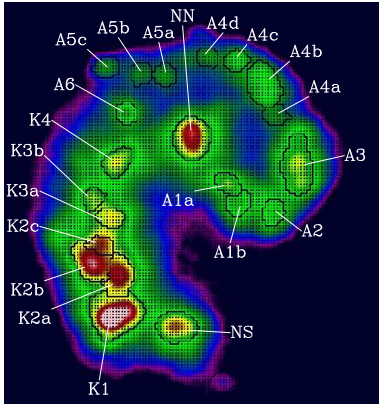}
   \caption{GROW$+$BORDER map example of the HIIphot program \citep{thilker00} 
            on the PACS 70\,$\mu$m map illustrating the photometric apertures 
            for the emission knots.}
   \label{HIIPhotexample}
   \end{center}
   \end{figure}

\begin{table}[h!]
\caption{Integral fluxes and fluxes of the emission knots from the ISOCAM 
15\,$\mu$m, {\em Spitzer}-MIPS 24\,$\mu$m, and {\em Herschel}-PACS 70, 100, and 160\,$\mu$m 
maps. The labelling of the knots follows Fig.~\ref{HIIPhotexample}. The 
photometric accuracy is estimated to be $\pm$20\%.}
\renewcommand{\footnoterule}{} 
\begin{center}
\begin{tabular*}{\columnwidth}{c@{\extracolsep\fill}c@{\extracolsep\fill}c@{\extracolsep\fill}c@{\extracolsep\fill}c@{\extracolsep\fill}c@{\extracolsep\fill}}
\hline
\hline
knot ID & \multicolumn{5}{c}{fluxes (mJy)} \\
        & 15\,$\mu$m & 24\,$\mu$m & 70\,$\mu$m & 100\,$\mu$m & 160\,$\mu$m \\
\hline
\hline
integral&    1720    &    5960    &   67200    &   102000    &   95500     \\
\hline
  NS    &    34.8    &    72.0    &    1726    &     1728    &    1489     \\
  K1    &   341.5    &    1282    &    8144    &     8567    &    5030     \\
  K2a   &    61.0    &   159.9    &   2529     &     4004    &  6894$^2$   \\
 K2b$+$c&    81.2    &   251.9    &   3474     &     5999    &             \\
  K3a   &    14.0    &    43.0    &   529.1    &   1549$^1$  &    ---      \\
  K3b   &     7.8    &    21.9    &   316.1    &             &    ---      \\
  K4    &    20.2    &    39.9    &   742.6    &    847.5    &   509.1     \\
  NN    &    84.7    &   260.7    &    3360    &     4904    &    4470     \\
 A1a$+$b&    17.5    &    34.0    &   708.2    &    858.4    &   500.7     \\
  A2    &     6.0    &    17.9    &   282.4    &    508.0    &    ---      \\
  A3    &    48.0    &   116.4    &    1695    &     1699    &    1353     \\
  A4b   &    12.1    &    20.5    &   698.6    &    693.8    &   744.6$^3$ \\
  A4c   &     5.3    &    19.9    &   216.3    &    368.8    &             \\
  A4d   &     3.8    &     5.9    &    81.6    &     ---     &    ---      \\
  A5a   &     3.3    &     5.7    &    64.9    &   340.2$^1$ &  441.1$^2$  \\
  A5b   &     3.7    &     4.5    &    66.5    &             &             \\
  A5c   &     2.7    &     4.9    &   110.5    &    237.9    &             \\
  A6    &     9.2    &    26.8    &   269.9    &    318.6    &    ---      \\
\hline
\multicolumn{6}{l}{\footnotesize $^1$ a$+$b~~~~~$^2$ a$+$b$+$c~~~~~$^3$ b$+$c} \\
\end{tabular*}
\end{center}
\label{hiiphotresults}
\end{table}

The two nuclei and the chain of HII region complexes along the northwestern 
wound-up spiral arm of NGC\,4038 are clearly resolved, as are several 
emission knots along an arc in the overlap region connecting the two nuclei
(see Fig.~\ref{HIIPhotexample} for the knot identification). 
Both nuclei are brighter than the HII regions in the spiral arm,
but the brightest emission in all three bands comes from the overlap region.
At 100 $\mu$m the brightest emission (K1 east) lies at the southern edge of the
overlap region. It coincides with the  prominent 15 and 24 $\mu$m emission 
peaks \citep{vigroux96,mirabel98}, which is located at the 
border of an optically bright HII region complex toward the dark dust lanes
(Fig.~\ref{pacs100contoursonHSTBVHalpha}). This region houses the supergiant 
molecular cloud complexes SGMC3-5 on the CO (1-0) map by \citet{wilson00}.
Two more bright FIR emission knots (K2a, K2b) are located north of K1
towards the optically obscured part of the overlap region 
(Fig.~\ref{pacs100contoursonHSTBVHalpha}) and coincide with SGMC1 and
SGMC2 on the CO (1-0) map by \citet{wilson00}, peaks number 3 and 4 on the
radio maps by \citet{hummel86} and knot 2 of the 850 $\mu$m map
by \citet{haas00}. Knots K2a and K2b together become the brightest area
in the 160\,$\mu$m map.

   \begin{figure}[ht!]
   \begin{center}
   \includegraphics[width=0.7\columnwidth,angle=90]{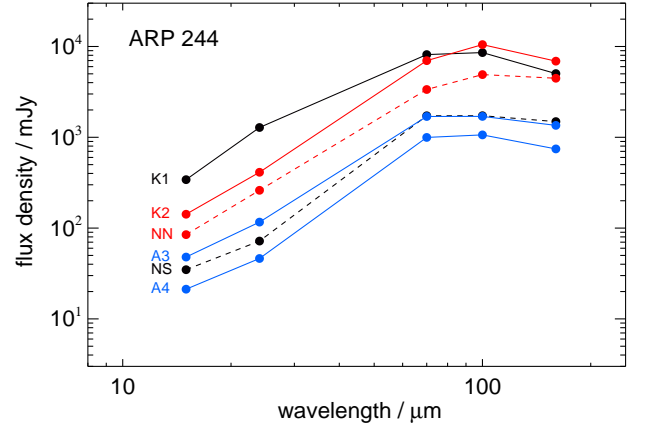}
   \includegraphics[width=0.7\columnwidth,angle=90]{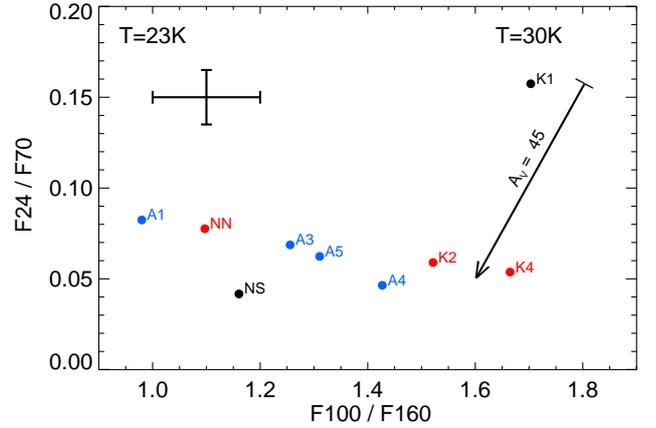}
   \caption{top: Inter-comparison of the SEDs of the emission knots.
            bottom: 2 colour diagram of the emission knots. Uncertainties
            due to the photometric accuracies are indicated by the cross.
            The 100/160 colour temperature range corresponding to a modified BB
            of shape $\nu^2 \times$ B($\nu$, T) and the
            extinction vector for A$_{\rm V}$ = 45\,mag are indicated. 
            The extinction curve of \citet{mathis83} has been used.}
   \label{knotanalysis}
   \end{center}
   \end{figure}

Table~\ref{hiiphotresults} presents the HIIphot \citep{thilker00} photometry
of the whole system and the individual emission knots as labelled in 
Fig.~\ref{HIIPhotexample}. The integral photometry is very consistent with
IRAS \citep{sanders03} and ISOPHOT \citep{klaas97} photometry. 
It is obvious that knots K1 and K2 by far exceed the brightness of the two
nuclei NS(4039) and NN(4038), also in the FIR. They are about 10 times brighter 
than the knots on the spiral arm.

In Fig.~\ref{knotanalysis} we compare the MIR and FIR photometric 
properties of the most prominent and a few selected, less bright knots.
All SEDs peak between 70 and 100\,$\mu$m, with the exception of NN, 
which is among the coldest emission knots. K1 clearly dominates the spectrum
up to 70\,$\mu$m, but for the longer wavelengths, it is slightly overtaken by 
the K2 area. From the colour-colour diagram in Fig.~\ref{knotanalysis} it 
becomes obvious that the overlap area houses the most active heating sources. 
K1, K2, and K4 all have dust colour temperatures above 27\,K, while the arm 
regions and the two nuclei are around 24\,K. K1 sticks out owing to its 
24\,$\mu$m excess. It is remarkable that K1 gets an SED comparable to K2 when 
reddened with a visual extinction of A$_{\rm V}$ = 45\,mag. However, none of 
the other knots including the ones from the spiral arm reaches such a high 
24/70 ratio as K1.  

\begin{table*}[ht!]
\caption{Derived properties of the FIR emission knots. For the meaning of the 
columns, see text. Values in parentheses are uncertain.}
\renewcommand{\footnoterule}{} 
\begin{center}
\begin{tabular*}{\textwidth}{c@{\extracolsep\fill}c@{\extracolsep\fill}c@{\extracolsep\fill}c@{\extracolsep\fill}c@{\extracolsep\fill}c@{\extracolsep\fill}c@{\extracolsep\fill}c@{\extracolsep\fill}c@{\extracolsep\fill}c@{\extracolsep\fill}c@{\extracolsep\fill}}
\hline
\hline
knot ID & size & L$_{\rm fir}$ & T$_{\rm 100/160}$ & M$_{\rm dust}$ & M$_{\rm H_{2}}$ & SFR & log$\sum$SFR & SFE & F$_{\rm 6cm}$ & log(F$_{\rm 6cm}$/F$_{\rm 70\mu,m}$)\\
        & (kpc$^2$) & (10$^{9}$ L$_{\odot}$) & (K) &  (10$^{6}$ M$_{\odot}$) &  (10$^{8}$ M$_{\odot}$) & (M$_{\odot}$/yr) & (M$_{\odot}$\,yr$^{-1}$\,kpc$^{-2}$) &  (L$_{\odot}$\,M$_{\odot}^{-1}$) & (mJy) & \\
\hline
\hline
integral&      &128.6&  23.4 &296.2 & 118.8& 22.2 &       &      & 195& -2.54 \\
\hline
NS      & 1.67 & 2.9 &  24.4 & 7.56 & 11.4 & 0.50 & -0.55 & 2.53 &  8 & -2.33 \\
K1      & 4.68 &16.1 &  30.0 & 3.63 & 17.0 & 2.78 & -0.23 & 9.47 & 16 & -2.71 \\
K2a     & 1.99 & 4.6 &  27.5 & 1.32 & 13.8 & 0.79 & -0.40 & 3.33 & 16 & -2.20 \\
K2b$+$c & 4.61 & 6.2 &  27.3 & 2.00 & 21.6 & 1.08 & -0.63 & 2.88 &  9 & -2.59 \\
K4      & 1.50 & 1.1 &  29.5 & 0.20 &  --  & 0.19 & -0.90 &  --  & (2)&(-2.57)\\
NN      & 3.22 & 6.8 &  23.7 & 4.79 & 31.0 & 1.17 & -0.44 & 2.19 & 12 & -2.45 \\
A3      & 2.69 & 2.9 &  25.4 & 0.24 &  3.3 & 0.50 & -0.73 & 8.78 & 10 & -2.23 \\
\hline
\end{tabular*}
\end{center}
\label{knotproperties}
\end{table*}

In Table~\ref{knotproperties} we have compiled important physical parameters 
of the emission knots: Size, corresponding to n$_{\rm pix}$ of the HIIphot aperture
(col.\ 2), far-infrared luminosity, derived from multi-component modified BB fits 
to the photometry of Table~\ref{hiiphotresults} (col.\ 3), colour temperature for 
a modified BB with $\beta$=2 (col.\ 4), dust mass, as described in \citet{wilke04} 
(col.\ 5), molecular gas mass, adding up all GMCs listed by \citet{wilson03} in the 
HIIphot aperture and re-scaling the distance (col.\ 6), star formation rate and 
star formation density according to \citet{kennicutt98} (cols.\ 7 \& 8), star 
formation efficiency as the ratio of far-infrared luminosity and molecular gas mass 
(col.\ 9), 6\,cm radio flux from \citet{hummel86} (col.\ 10), and radio to 
far-infrared flux ratio according to \citet{dejong85} (col.\ 11).
The luminosity distance D = 28.4\,Mpc (from NED) was used for all 
distance-dependent measures.

\section{Discussion}

The Antennae is an interacting luminous (10$^{11}$ L$_{\rm \odot}$) IR system close
to the 2nd encounter when the two galaxy disks will start to merge into a single
system \citep[e.g.][]{karl08}. The new FIR observations, along with the available 
radio and CO data, suggest that the gravitational forces have shuffled a huge 
amount of interstellar material along an arc connecting the two galaxy nuclei. 
Several kpc-sized, large ISM complexes house the most active star formation sites 
and evolve independently. With respect to their position on the optical image 
(Fig.~\ref{pacs100contoursonHSTBVHalpha}), knots K2 and K3 appear to be located 
directly at the collision front of the two disks embedded in dense dust lanes but
already associated with bright star clusters. Knot K1 is located more in
the disk of NGC4039, while knot K4 is located in the disk of NGC\,4038.

Knot K4 is a star formation area burning on a relatively low level. No
CO emission is detected and there is no noticeable dust lane, so that it appears 
to have consumed its fuel and will start to extinguish soon.

Knot K2 reaches 2/3 of the luminosity of K1, however the derived star formation 
densities and efficiencies are considerably lower than for K1. The spatially 
resolved SEDs obtained from the PACS maps clarify that it is more active than was 
suggested by \citet{haas00} on the basis of the ISO and the 450 and 850 $\mu$m 
SCUBA maps. If K2 suffered from higher obscuration than K1 (see 
Fig.~\ref{knotanalysis}), the intrinsic difference between K2 and K1 would become 
smaller. There seems, however, to be a dichotomy between K2a and K2b$+$c. While 
the latter appears to have a normal radio-to-FIR ratio, K2a shows an excess in 
radio emission, as already noted by \citet{chyzy04}. It is very inconspicuous in 
the optical and is lacking X-ray emission \citep[see][Fig. 9]{fabbiano01}. K2a 
might therefore be in a younger evolutionary stage than K2b$+$c.

Knot K1 is currently the most active area by far. This is manifested by the
highest individual luminosity, which itself is as high as the Milky Way
luminosity but concentrated in an area of 2\,kpc diameter. It exhibits
the highest star formation rate and efficiency, and the highest star 
formation density that consequently raises dust temperatures to a maximum. The 
high density of hot heating sources can explain the very high MIR emission. Its 
radio-to-FIR luminosity ratio deviates from the average in that there is a lack 
of radio emission, and at the peak position, there is no X-ray emission either 
\citep[see][Fig. 9]{fabbiano01}. These findings classify it as a very young complex.

\section{Conclusions}

New {\em Herschel}-PACS scan maps of the Antennae at 70, 100, and 160\,$\mu$m provide
a high spatial resolution complement to earlier MIR maps (ISOCAM, {\em Spitzer}-MIPS),
high-resolution CO (1-0), and radio 6\,cm maps and thus allow detailed study of 
the star formation state in individual emission knots. 

We confirm that the highest star formation activity takes place in the overlap
area concentrated in the two emission complexes K1 and K2. Our star formation 
diagnostics strongly indicates that these knots differ in evolutionary stage. 

\begin{acknowledgements}
      PACS has been developed by a consortium of institutes led by 
      MPE (Germany) and including UVIE (Austria); KUL, CSL, IMEC (Belgium); 
      CEA, OAMP (France); MPIA (Germany); IFSI, OAP/AOT, OAA/CAISMI, LENS, 
      SISSA (Italy); IAC (Spain). This development has been supported by the 
      funding agencies BMVIT (Austria), ESA-PRODEX (Belgium), 
      CEA/CNES (France), DLR (Germany), ASI (Italy), and CICYT/MCYT (Spain).
      M.H. was supported by the Nordrhein-Westf\"alische Akademie der 
      Wissenschaften und der K\"unste. 
      This research made use of the NASA/IPAC Extragalactic Database (NED).
      HST images were retrieved from the Canadian Astronomical Data Centre (CADC)
      Virtual Observatory via the Aladin tool. We thank the referee for 
      constructive comments.
\end{acknowledgements}

\bibliographystyle{aa}
\bibliography{14670.bib}

\end{document}